\documentclass[12pt]{article}

\newcommand{\beq}{\begin{equation}}
\newcommand{\eeq}{\end{equation}}
\newcommand{\beqn}{\begin{eqnarray}}
\newcommand{\eeqn}[1]{\label{#1}\end{eqnarray}}

\newcommand{\id}
{i\kern.06em\hbox{\raise.25ex\hbox{$/$}\kern-.60em$\partial$}}

\newcommand{\bs}{/\kern-.52em b}
\newcommand{\qs}{/\kern-.52em s}

\newcommand{\dd}
{\kern.06em\hbox{\raise.25ex\hbox{$/$}\kern-.60em$\partial$}}

\newcommand{\tr}{\mathop{\rm tr}\nolimits}

\begin{document}

\begin{titlepage}
\title{Remarks on Goldstone bosons and hard thermal loops}

\author{
G. Alexanian$^a$\thanks{E-mail address:
garnik@scisun.sci.ccny.cuny.edu}\, ,
E. F. Moreno$^{a,b}$\thanks{
Investigador CONICET, E-mail address:
moreno@scisun.sci.ccny.cuny.edu}\, ,
V. P. Nair$^a$\thanks{E-mail address:
vpn@ajanta.sci.ccny.cuny.edu}\,
and\,
R. Ray$^a$\thanks{E-mail address:
rray@scisun.sci.ccny.cuny.edu}
\\
\\
{\small\it $^a$ Physics Department,
City College of the City University of New York}\\
{\small\it New York NY 10031, USA}\\
{\small\it $^b$ Departamento de F\'\i sica,
Universidad Nacional de La Plata}\\
{\small\it C.C. 67, 1900 La Plata, Argentina}\\
}


\maketitle


\begin{abstract}
The hard thermal loop effective action for Goldstone bosons is
deduced by symmetry arguments from the corresponding result for
gauge bosons. Pseudoscalar mesons in Chromodynamics and magnons
in an antiferromagnet are discussed as special cases, including
the hard thermal loop contribution to their scattering.
\end{abstract}
\end{titlepage}


The importance of hard thermal loops (HTL's) in a thermal gauge
theory was recognized a few years ago \cite{BP}. The proper
identification of the HTL-contributions and the resummation of
Feynman diagrams to take into account their effects are a crucial
first step towards a thermal perturbation theory for gauge
fields, which is free of infrared singularities. The
HTL-contributions in a gauge theory can be summarized by an
effective action, different versions of which have been analyzed
in detail by various groups \cite{BP,FT,EN,BI}. More recently, it
has been pointed out that there are HTL-contributions in the
chiral model for pions or more generally in a theory of Goldstone
bosons \cite{PT,Man}. Since Goldstone bosons behave in a way
similar to the longitudinal polarizations of massive gauge
bosons, we can expect that the HTL's for Goldstone bosons should
be related to the HTL's for gauge bosons via symmetry arguments.
Some elements of this connection are evident in references
\cite{PT,Man}. Nevertheless the arguments presented there are not
entirely symmetry-based. It should be possible to deduce the HTL
effective action for Goldstone bosons purely by symmetry
arguments starting from the HTL action for gauge bosons. In this
note, we present the relevant arguments, for Goldstone bosons
corresponding to a global symmetry group $G$ being spontaneously
broken to $H \subset G$. The basic strategy is to rewrite the
dynamics of the Goldstone bosons as a gauge theory with gauge
group $H$ and then to use this gauge theory result with
appropriate minor changes. As special cases, we consider
$G=SU_L(N_f)\times SU_R(N_f)\, , H=SU_{L+R}(N_f)$ corresponding
to the pseudo scalar mesons and $G=SU(2)\, , H=U(1)$
corresponding to magnons or spin waves in an antiferromagnet.

The Goldstone boson fields corresponding to the symmetry breaking
$G \to H$ take values in the coset $G/H$ and their dynamics can
be described by a nonlinear sigma model with target space $G/H$.
We begin with a brief description of this theory as a theory with
$H$-gauge symmetry \cite{BST}. Let $T^{\alpha} , \; \;
\alpha=1,\cdots,dim\; G$ denote the generators of $\cal G$ and
$t^a , \; \; a=1,\cdots,dim\; H$ denote the generators of ${\cal
H}$. We assume the standard normalization $Tr(T^{\alpha}
T^{\beta}) = 1/2\; \delta^{\alpha \beta}$, for the fundamental
representation of the generators. The generators in the
orthogonal complement of $\cal H$ in $\cal G$ will be denoted by
$S^i , \; \; i=1,\cdots, dim\; G - dim\; H$. The commutation
rules are of the form
\begin{eqnarray}
&&[t^a, t^b] = i f^{a b c} t^c\; , \;\;\;\;\;\;
[t^a, S^i] = i \left(D^a\right)^{i j} S^j \nonumber\\
&&[S^i, S^j] = i f^{a i j} t^a .
\label{s-1}
\end{eqnarray}

The structure of these commutation rules, with $[S,S]\approx t$
implies that we are considering the case when $G/H$ is a
symmetric space. Let $g(x)$ be a $G$- valued field. Define
\begin{equation}
V_{\mu}^a = 2\; Tr\left(t^a \partial_{\mu} g g^{-1}\right)\; ,
\;\;\;\;\;
E_{\mu}^i = 2\; Tr\left(S^i \partial_{\mu} g g^{-1}\right) .
\label{s-2}
\end{equation}
This corresponds to the decomposition $\partial_{\mu} g g^{-1} =
V_{\mu} + E_{\mu} , \;\; V_{\mu} = t^a V_{\mu}^a , \;\; E_{\mu} =
S^i E_{\mu}^i$. Under $H$- transformations of $G$ on the left,
{\it i.e.,} under $g \to g'= h g$, $V_{\mu}$ transforms as a
gauge potential, namely,
\begin{equation}
V_{\mu}(h g) = h V_{\mu} h^{-1} + \partial_{\mu} h h^{-1} .
\label{s-3}
\end{equation}
The field strength associated with this gauge potential is given
by
\begin{eqnarray}
F_{\mu \nu} &=& \partial_{\mu} V_{\nu} - \partial_{\nu} V_{\mu} -
[V_{\mu}, V_{\nu}]\nonumber\\
&=& \left( -i t^a\right) f^{ a i j} E^i_{\mu} E^j_{\nu} .
\label{s-4}
\end{eqnarray}
The gauge potential $V_{\mu}$ also allows us to the define the
covariant derivative
\begin{equation}
D_{\mu} g = \partial_{\mu} g - V_{\mu} g .
\label{s-5}
\end{equation}
The Lagrangian for the $G/H$- sigma model may be written as
\begin{equation}
L= -\alpha Tr\left( D_{\mu} g g^{-1}\;  D^{\mu} g g^{-1} \right)
.
\label{s-6}
\end{equation}
This Lagrangian has invariance under the global $G$
transformations $g\to g\; U\; , \;\; U\in G$, as expected for a
theory for which the symmetry breaking $G\to H$ is only
spontaneous. Further, it has invariance under the local $H$-gauge
transformations $g(x) \to h(x) g(x)$. The field $g(x)$ has $dim\;
G$ degrees of freedom. The $H$-gauge invariant shows that it is
possible to ``gauge away" the degrees of freedom corresponding to
$H$, leaving only $G/H$-degrees of freedom. (This can general be
done only locally in some parametrizations of $g$ and $H$, since,
in general, $G\neq G/H \times H$.) With the splitting
$\partial_{\mu} g g^{-1} = V_{\mu} + E_{\mu}$, we find $L = -
\alpha/2 \; E_{\mu}^i E^{i \mu}$, which is proportional to the
Cartan-Killing metric on the coset space $G/H$. Thus (\ref{s-6})
is indeed equivalent to the standard sigma model for $G/H$.

The Lagrangian (\ref{s-6}) describes the $G/H$-model as a theory
of ``matter fields" minimally coupled to an $H$-gauge potential
$V_{\mu}$. At finite temperature, therefore we expect a hard
thermal loop mass term for the gauge field $V_{\mu}$, due to the
electrical screening effects of the matter fields in $G/H$. Now,
the HTL-effective action for a pure gauge theory with no matter
fields, is given in terms of the gauge potential $A_{\mu}$ as
\cite{EN}
\begin{equation}
\Gamma[A]_{gauge} = \frac{C_G T^2}{6} \int d\Omega\; d^2x^T\;
S_{WZW} (N^{-1} M)
\label{s-7}
\end{equation}
where $C_G$ is the quadratic Casimir for the adjoint
representation of the group and $S_{WZW}$ is the
Wess-Zumino-Witten action defined on the two-dimensional space of
$x^{\pm} = 1/2\;(x^0\mp{\vec Q}\cdot{\vec x})$. i.e.,
\begin{eqnarray}
S_{WZW}(U)&=&{1\over2\pi}\int_Mdx^+ dx^-~\tr(\partial_+U\partial_{-}
U^{-1})\nonumber\\
&~&~~~-{i\over 12\pi}
\int_{M^3}d^3x~\epsilon^{\mu\nu\alpha}\tr(U^{-1}\partial_\mu U
U^{-1}\partial_\nu U U^{-1}\partial_\alpha U)
\label{WZW}
\end{eqnarray}
$M, N$ are defined by $A_{+}=1/2\; (A_0 + {\vec Q} \cdot {\vec
A}) = -\partial_{+} M M^{-1}$, $A_{-}=1/2\; (A_0 - {\vec Q} \cdot
{\vec A}) = -\partial_{-} N N^{-1}$. $d\Omega$ denotes
integration over the orientation of the unit vector ${\vec Q}$;
integration over coordinates transverse to ${\vec Q}$, {\it
viz.,}$x^T$, is explicitly shown in (\ref{s-7}) while integration
over $x^{\pm}$ is included in the definition of $S_{WZW}$.

For the $G/H$-model, the result should be similar to (\ref{s-7})
with $A_{\mu}$ replaced by $V_{\mu} = t^a\; 2 Tr\left( t^a
\partial_{\mu}g g^{-1}\right)$. The overall coefficient will be
different. In the case of gauge bosons, there are two
polarization states which contribute to the screening; for
Goldstone bosons we have only one polarization state. This should
give an additional factor of $1/2$. Further, for the $G/H$ model,
the $V_{\mu}$-fields couple only to the $G/H$-degrees of freedom,
the coupling charge matrices being $f^{a i j}$ from (\ref{s-1}).
Since $f^{a i j} f^{b i j}=f^{a \alpha \beta} f^{b \alpha \beta}
- f^{a c d} f^{b c d} = (C_G-C_H)\; \delta^{a b}$, we see that
$C_G$ in eq.(\ref{s-1} should be replaced by $C_G-C_H$. The
HTL-effective action for the Goldstone modes in $G/H$ can thus be
written as
\begin{eqnarray}
\Gamma[V] &=& \frac{T^2}{12} \left(C_G-C_H\right)\; \int d\Omega
d^2 x^{T}\; S_{WZW}(N^{-1} M)\nonumber\\
&=& \frac{1}{2} \left. \frac{C_G-C_H}{C_G}\;
\Gamma[A]\right|_{A_{\mu} \to V_{\mu}} .
\label{s-8}
\end{eqnarray}

This result has been obtained purely by symmetry arguments. It
can be checked by explicit calculations or by comparison to
previous calculations as we shall do shortly.

Notice that $\Gamma$ as given by (\ref{s-8}), is at least quartic
in the Goldstone fields. Since $\Gamma$ is gauge-invariant, the
$H$-degrees of freedom can be removed; by orthogonality of $t^a$
and $S^i$ and the commutation rules (\ref{s-1}), up to an H-gauge
transformation, $V_{\mu}$ is at least quadratic in the Goldstone
fields:
\begin{eqnarray}
V_{\mu}^a &=& 2 Tr\left( t^a\; \partial_{\mu}e^{i \pi^i S^i}\;
e^{-i \pi^i S^i}\right) \nonumber\\
&&\hspace{-1.0cm}\approx 2 Tr\; t^a (i\partial_{\mu}
\pi^i S^i + \partial_{\mu} \pi^i \pi^j [S^i,S^j] + \cdots)
 \approx i f^{a i j} \partial_{\mu}\pi^i \pi^j + \cdots
\end{eqnarray}
$\Gamma$ being quadratic in $V_{\mu}$'s, the lowest order
term in (\ref{s-8}) is quartic in the Goldstone fields.

A comment regarding the direct evaluation of the result in terms
of the Goldstone fields is in order. In terms of the gauge field
$V_\mu$, the leading term in (\ref{s-8}) is quadratic and this
can be evaluated by the two-point vacuum polarization diagram
with $V_\mu$ on the external lines. A comparison of the overall
coefficient in (\ref{s-8}) can thus be done with the explicit
evaluation of the vacuum polarization diagram. However, for the
term with four external Goldstone particles, higher diagrams with
upto four external lines can in principle contribute. Directly in
terms of Goldstone fields, the orders of various terms can get
mixed up, since $V_\mu$ is itself made of the Goldstone fields
and obeys identities like (\ref{s-4}) ( where the curl of $V_\mu$
is related to a term quadratic in the fields). In seeking a
covariant generalization of the result of the vacuum polarization
diagram, this point must be taken care of. One must keep $V_\mu$
as an arbitrary external field and compare the coefficient of
(\ref{s-8}) with the evaluation of the vacuum polarization
diagram. This seems to account for the discrepancy of a factor of
4 between references \cite{PT} and \cite{Man}.

The result for pions given in references \cite{PT,Man} also
include the leading $T^2$-correction to the coefficient $\alpha$
in the chiral Lagrangian (\ref{s-6}). Such a correction, which
can contribute at the quadratic order in the  Goldstone fields,
is not, from our point of view, a hard thermal loop contribution.
To see how this arises, consider a background field expansion of
(\ref{s-6}). Writing $g=U\; B$, where $B$ denotes the background
field, and $U=exp\left(i\varphi^j S^j\right)$ we find
\begin{equation}
L=\frac{1}{2} \left(D_{\mu} \varphi\right)^2 + 2 {\cal A}_{\mu}^i
{\cal A}_{\mu}^i + 2 \varphi^j \varphi^k f^{j m l} f^{k n l}
{\cal A}_{\mu}^m {\cal A}_{\mu}^n + \cdots \label{s-9}
\end{equation}
where ${\cal A}_{\mu}^i=1/2\; \left(\partial_{\mu} B
B^{-1}\right)^i$, $D_{\mu}^{i j}  = \partial_{\mu} \delta^{i j} +
f^{i j a} V_{\mu}^a$, $V_{\mu}^a =1/2\; \left(\partial_{\mu} B
B^{-1}\right)^a$. The first  term shows the $H$-gauge invariant
structure and leads to the result (\ref{s-8}) as we have argued.
The last term gives, upon Wick contraction of $\varphi$'s with a
thermal propagator,
\begin{equation}
\delta \Gamma = - 2 \left[\frac{T^2}{24}
(C_G-C_H)\right] \int\; {\cal A}^2
\label{s-10}
\end{equation}
which corresponds to the modification $\alpha\to \alpha(T)$,
\begin{equation}
\alpha(T)= \alpha - \frac{T^2}{24} (C_G-C_H) .
\label{s-11}
\end{equation}
To leading order in $T^2$ and in HTL-approximation, (\ref{s-8})
and (\ref{s-10}) are the only corrections.

We now consider the specialization of the results (\ref{s-8}),
(\ref{s-10}) to the case of pions or pseudoscalar mesons. In this
case $G=SU_L(N_f)\times SU_R(N_f)\, , H=SU_{L+R}(N_f)$. $G$ may
be parametrized by $(g_1, g_2)\; , \;\; g_i(x)\in SU(N_f)$. The
gauge potential is given by$V_{\mu}=1/2\; \left(\partial_{\mu}
g_1 g_1^{-1} - g_2^{-1} \partial_{\mu} g_2\right)$ with $H$
transformations acting as $g_1\to h(x) g_1\; ,\;\; g_2\to g_2
h^{-1}(x)\; , \;\; h(x)\in SU(N_f)$. Global $G$-transformations
act as $g_1\to g_1 U_L\; ,\;\; g_2\to U_R g_2\; , \;\; U_L,U_R\in
G$. The Lagrangian (\ref{s-6}) becomes
\begin{eqnarray}
L&=&-\alpha\; Tr\left[\left(g_1^{-1} D_{\mu}g_1\right)^2 +
\left(g_2 D_{\mu} g_2^{-1}\right)^2\right]\nonumber\\
&=& - 2 \alpha Tr\left( {\cal A}_{\mu}^2\right)
\label{s-12}
\end{eqnarray}
where $D_{\mu}=\partial_{\mu} - V_{\mu}$ and ${\cal A}_{\mu} =
1/2\; \left(\partial_{\mu} g_1 g_1^{-1} + g_2^{-1} \partial_{\mu}
g_2\right)$. The $H$-symmetry allows us to chose a gauge where
$g_2=1$ or equivalently we can consider $g_2 g_1=U(x)\in SU(N_f)$
as the residual degrees of freedom. In this gauge $V_{\mu}={\cal
A}_{\mu}=1/2\; \left(\partial_{\mu}U U^{-1}\right)$ and
$L=-\alpha/2 \; Tr\left(\partial_{\mu}U U^{-1}\right)^2$ which is
the usual chiral Lagrangian with $\alpha=2 f_{\pi}^2$. In this
case, by expansion of (\ref{s-8}) in powers of $V_{\mu}$, we can
check by direct comparison that (\ref{s-8}) agrees with the
result of references \cite{PT,Man}. Furthermore, from
(\ref{s-11}),
\begin{equation}
f_{\pi}^2(T) = f_{\pi}^2 - \frac{N_f T^2}{48}
\label{s-13}
\end{equation}
which also agrees with the result in references \cite{PT,Man},
noting that with our normalization for the generators, our
$f_{\pi}^2$ is $1/4$ of the $f_{\pi}^2$ used in \cite{PT,Man}.

Using equation (\ref{s-8}) we can evaluate the pion-pion
scattering amplitude for the process $(E_1,{\vec
k}_1,e^1),(E_2,{\vec k}_2,e^2) \rightarrow (E_3,{\vec
k}_3,e^3),(E_4,{\vec k}_4,e^4)$, ($e^1, e^2, e^3$, and $e^4$ are
polarization vectors), where the pion fields are related to the
field $U$ through the identity $U=exp\; ({i \pi^i t^i/f_{\pi}})$
(we are considering here the case $N_f=2$).The result can be
computed to be:
\begin{eqnarray}
{\cal A}&=& {{i\; \delta^4 (k_1+k_2-k_3-k_4)}\over (2 \pi)^2\;
\prod_i \sqrt{2\;E_i}}~{\cal M} ,\nonumber\\
{\cal M}&=& A\; (e^1\cdot e^2)\; (e^3\cdot e^4) ~+~
B\; (e^1\cdot e^3)\; (e^2\cdot e^4) ~+~\nonumber\\
&& \hspace{1cm} C\; (e^1\cdot e^4)\; (e^2\cdot e^3) \; ,
\nonumber\\
A&=&\frac{1}{4 f_{\pi}^2(T)} \left(k_1\cdot k_2 + k_3\cdot k_4 \right)
- \frac{T^2}{192 f_{\pi}^4}  \left[ \left(k_1+k_3\right)_{\mu}
M_{\mu \nu}(k_1-k_3)\times \right. \nonumber\\
&& \left.\left(k_2+k_4\right)_{\nu} +
\left(k_1+k_4\right)_{\mu} M_{\mu \nu}(k_1-k_4)
\left(k_2+k_3\right)_{\nu}\right]\; ,\nonumber\\
B&=&-\frac{1}{4 f_{\pi}^2(T)} \left(k_1\cdot k_3 + k_2\cdot k_4 \right)
+ \frac{T^2}{192 f_{\pi}^4}  \left[ \left(k_1-k_2\right)_{\mu}
M_{\mu \nu}(k_1+k_2)\times \right. \nonumber\\
&& \left.\left(k_3-k_4\right)_{\nu} +
\left(k_1+k_4\right)_{\mu} M_{\mu \nu}(k_1-k_4)
\left(k_2+k_3\right)_{\nu}\right]\; ,\nonumber\\
C&=&-\frac{1}{4 f_{\pi}^2(T)} \left(k_1\cdot k_4 + k_2\cdot k_3 \right)
- \frac{T^2}{192 f_{\pi}^4}  \left[ \left(k_1-k_2\right)_{\mu}
M_{\mu \nu}(k_1+k_2)\times \right. \nonumber\\
&& \left.\left(k_3-k_4\right)_{\nu} -
\left(k_1+k_3\right)_{\mu} M_{\mu \nu}(k_1-k_3)
\left(k_2+k_4\right)_{\nu}\right]\; .
\label{scatt-pions}
\end{eqnarray}
The bilinear kernel $M_{\mu \nu} (p)$ is given by
\begin{equation}
M_{\mu \nu}(p)=g_{\mu 0}\; g_{\nu 0} - p^0 \int \frac{d
\Omega_Q}{4 \pi}\; \frac{Q_{\mu} Q_{\nu}}{p\cdot Q}
\label{m}
\end{equation}
(here $Q$ is the null vector $(1,{\vec q})\;,\;\; {\vec q}^2=1$).

The expression (\ref{scatt-pions}) takes a particularly simple
form if the total (spatial) momentum is zero: ${\vec k}_1+{\vec
k}_2=0$, $E_i\equiv E=|{\vec k}_1|$ and the scattering angle is
defined by ${\vec k}_1 \cdot {\vec k}_3 = |{\vec k}_1||{\vec
k}_3| \cos \theta$. Then
\begin{eqnarray}
A&=& {E^2\over f^2_{\pi}(T)} \left( 1-{T^2\over 24
f^2_{\pi}(T)}\right) \approx {E^2 \over f^2_{\pi}},\nonumber\\
B&=&B_1-B_2\; cos(\theta)\; , \;\;\;\;\;
C=B_1+B_2\; cos(\theta)\; ,\nonumber\\
B_1&=& -{E^2\over 2\;f^2_{\pi}(T)} \left( 1-{T^2\over
24\;f^2_{\pi}(T)}\right) \approx -{E^2 \over 2\;f^2_{\pi}},\nonumber\\
B_2&=& -{E^2\over 2\;f^2_{\pi}(T)} \left( 1-{T^2\over
72\;f^2_{\pi}(T)}\right) \approx -{E^2 \over 2\;f^2_{\pi}}\left(
1+{T^2\over 36 f^2_{\pi}} \right) .
\label{s-14b}
\end{eqnarray}
Notice that the contribution of the hard thermal loops is
comparable, and with opposite sign, to the other leading
$T$-dependent corrections. Moreover, for a scattering angle of
$\theta= \pm \pi/2$ the scattering amplitude is independent of
the temperature.

We now consider the case of spin waves or magnons in an
antiferromagnet \cite{burgess}. Since the dispersion relation is
linear for antiferromagnetic magnons (as opposed to the
ferromagnetic case), it is for this case that it is possible to
adapt equations (\ref{s-8}) and (\ref{s-10}) in a simple way. The
groups involved are $G=SU(2)$ and $H=U(1)$. A convenient
parametrization for $g\in SU(2)$ is
\begin{equation}
g=\lambda \left( \matrix{1&z\cr
-{\bar z}&1\cr}\right) {1\over \sqrt{1+z{\bar z}}}
\label{s-15}
\end{equation}
where $\lambda = \exp (i\sigma^3 \theta /2)\in U(1).~ (z,{\bar z})$
parametrize the coset $SU(2)/U(1)$. From $\partial_\mu g~g^{-1}$, we
identify
\begin{equation}
V_\mu = i{({\bar z}\partial_\mu z -\partial_\mu {\bar z} z)\
\over (1+z{\bar z})}
\label{s-16}
\end{equation}

Specialization of (\ref{s-8}) to the magnon case is obtained by
taking $G=SU(2), H=U(1)$ and $A^{1,2}_\mu =0, A^3_\mu = V_\mu$.
In addition, we have to incorporate the fact that magnons have a
propagation speed $v$ which is not 1. The dispersion relation
$\omega =v \vert {\vec k}\vert$ shows that every spatial
derivative should carry a factor of $v$. In other words, we need
$\partial_\mu \rightarrow {\tilde\partial}_\mu =(\partial_0,
v\partial_i)$. Further there must be a factor of $(1/ v^3)$ in
$\Gamma$ for dimensional reasons. This can also be seen
diagrammatically as arising from $d^3k =k^2 dk d\Omega =  (1/v^3)
\omega^2 d\omega d\Omega$. Putting all this together
\begin{equation}
\Gamma = -{T^2 \over 24\pi v^3} \int {d^4k\over (2\pi)^4}
\left({{\bar z}{\tilde\partial}_\mu z - z {\tilde\partial}_\mu {\bar z}
\over 1+z{\bar z}}
\right) (-k) M_{\mu\nu}({\tilde k}) \left({\bar z}{\tilde\partial}_\mu
z -z{\tilde\partial}_\mu {\bar z} \over 1+z{\bar z}\right)(k)
\label{s-17}
\end{equation}
where $M_{\mu\nu}$ is given in equation (\ref{m}).

The kinetic energy term or the sigma model part of the action
is given by (\ref{s-6}) with appropriate changes as
\begin{equation}
S_0 = 2\alpha { {\tilde\partial}_\mu z {\tilde\partial}_\mu{\bar z}
\over (1+z{\bar z})^2} = {1\over 2} { {\tilde \partial}_\mu
\varphi_i {\tilde\partial}_\mu \varphi_i \over (1+{\varphi_i
\varphi_i \over 4\alpha})^2 }
\label{s-19}
\end{equation}
where $2\sqrt{\alpha} ~z= (\varphi_1-i\varphi_2)$ and
$\alpha (T) =\alpha (0) -(T^2/ 12v^3) $.

The hard thermal loop contribution is at least quartic in the
magnon fields and so can contribute to a $T$-dependent term to
magnon-magnon scattering. The quartic term in (\ref{s-19}) also
contributes to such a process. The magnon wave function can be
taken to be
\begin{equation}
\varphi_i = e^{(\lambda )}_i {\exp (-i(\omega t -{\vec k}\cdot
{\vec x}))\over \sqrt{2\omega V}}
\label{s-20}
\end{equation}
where $e^{(\lambda )}_i$ is the polarization and we choose
normalization in a volume $V$. Consider the scattering process
$(k_1,e^1),(k_2,e^2) \rightarrow (k_3,e^3),(k_4,e^4)$. The
amplitude for this process can be calculated to be

\begin{eqnarray}
{\cal A}&=& {{i(2\pi )^4 \delta (k_1+k_2-k_3-k_4)}\over
\prod_i \sqrt{2\omega_i V}}~{\cal M} ,\nonumber\\
{\cal M}&=& A\; (e^1\cdot e^2)\; (e^3\cdot e^4) ~+~
B\; (e^1\cdot e^3)\; (e^2\cdot e^4) ~+~\nonumber\\
&& \hspace{1cm} C\; (e^1\cdot e^4)\; (e^2\cdot e^3)
\nonumber\\
A&=&\frac{1}{\alpha(T)} \left(k_1\cdot k_2 + k_3\cdot k_4 \right)
- \frac{T^2}{48 \pi v^3 \alpha(T)^2}  \left[ \left(k_1+k_3\right)_{\mu}
M_{\mu \nu}(k_1-k_3)\times \right. \nonumber\\
&& \left.\left(k_2+k_4\right)_{\nu} +
\left(k_1+k_4\right)_{\mu} M_{\mu \nu}(k_1-k_4)
\left(k_2+k_3\right)_{\nu}\right]\nonumber\\
B&=&-\frac{1}{\alpha(T)} \left(k_1\cdot k_3 + k_2\cdot k_4 \right)
+ \frac{T^2}{48 \pi v^3 \alpha(T)^2}  \left[ \left(k_1-k_2\right)_{\mu}
M_{\mu \nu}(k_1+k_2)\times \right. \nonumber\\
&& \left.\left(k_3-k_4\right)_{\nu} +
\left(k_1+k_4\right)_{\mu} M_{\mu \nu}(k_1-k_4)
\left(k_2+k_3\right)_{\nu}\right]\nonumber\\
C&=&-\frac{1}{\alpha(T)} \left(k_1\cdot k_4 + k_2\cdot k_3 \right)
- \frac{T^2}{48 \pi v^3 \alpha(T)^2}  \left[ \left(k_1-k_2\right)_{\mu}
M_{\mu \nu}(k_1+k_2)\times \right. \nonumber\\
&& \left.\left(k_3-k_4\right)_{\nu} -
\left(k_1+k_3\right)_{\mu} M_{\mu \nu}(k_1-k_3)
\left(k_2+k_4\right)_{\nu}\right]\; .
\label{scatt-magnons}
\end{eqnarray}

Again, this expression is enormously reduced if the combined
momentum of the incoming magnons vanishes. In this case we have
\begin{eqnarray}
A&=& {4\omega^2\over \alpha (T)} \left( 1-{T^2\over 6
v^3\alpha(T)
}\right) \approx {4\omega^2 \over \alpha (0)} \left( 1-{T^2\over
12 v^3\alpha(0) }\right) ,\nonumber\\
B&=&B_1-B_2\; cos(\theta)\; , \;\;\;\;\;
C=B_1+B_2\; cos(\theta)\; ,\nonumber\\
B_1&=& -{2\omega^2\over \alpha (T)} \left( 1-{T^2\over 6v^3\alpha(T) }
\right)
\approx -{2\omega^2 \over \alpha (0)}\left( 1-{T^2\over 12v^3
\alpha(0)} \right) ,\nonumber\\
B_2&=& {2\omega^2\over \alpha (T)} \left( 1-{T^2\over 18v^3\alpha(T)}
\right) \approx {2\omega^2\over \alpha (0)} \left( 1+{T^2\over
36 v^3\alpha (0)}\right)\; .
\label{s-21b}
\end{eqnarray}
As in the case of pion scattering, the contribution of the hard
thermal loops is of the same order of magnitude as the other
leading $T$-dependent corrections. The temperature at which
$\alpha(T)$ vanishes, and thereby restores disorder, gives an
estimate of the N\'eel temperature $T_N$ as $T_N^2=12 v^3
{\alpha}(0)$. This is of course rather crude, the calculation of
$\alpha (T)$ cannot be trusted very near the transition point;
nevertheless it gives a rough estimate. The corrections to
scattering are thus seen to be proportional to $(T^2/T_N^2)$.

To recapitulate, we have shown in this article that the hard
thermal loop effective action for Goldstone bosons corresponding
to a symmetry breaking pattern $G \to H$ can be deduced entirely
by symmetry arguments. In particular we discuss two examples:
pseudoscalar mesons and magnons in an antiferromagnet. In both of
these cases we see that the Goldstone boson scattering amplitude
is modified significantly by the contribution from the hard
thermal loop term.

\section*{Acknowledgements}
We wish to thank C. Manuel for a critical reading of the manuscript.
G.A. and R.R. were supported in part by a PSC-CUNY grant. E.F.M.
was supported by CONICET. V.P.N. was supported in part by the
National Science Foundation grant number PHY-9605216.

\end{document}